# Orbital instability and the loss of quantum coherence

Allan Tameshtit and J. E. Sipe
*Department of Physics and Ontario Laser and Lightwave Research Centre,
University of Toronto, Toronto, Ontario, Canada M5S 1A7*

We compare quantum decoherence in generic regular and chaotic systems that interact with a thermal reservoir via a dipole coupling. Using a time-dependent, self-consistent approximation in the spirit of Hartree, we derive in the high temperature limit an expression for the off-diagonal elements of the system density operator that initially corresponds to a coherent superposition of two adjacent wave packets. We relate the decoherence rate to the Lyapunov exponent in the Ehrenfest regime. In this regime, the greater the instability of the system the faster the loss of coherence occurs.

## I. INTRODUCTION

Classical behavior is ubiquitous in a universe that is fundamentally quantum mechanical. From one point of view, this observation is rationalized by noting the similarity between the quantum and classical equations of Ehrenfest and Newton. But recent developments in the theory of chaos complicate this explanation. Realistic Hamiltonians give rise to significant regions of phase space that are classically chaotic. In such regions the Ehrenfest regime, in which quantum expectation values behave as do classical trajectories, is much more fleeting than in regular regions [1,2]. From this point of view, then, the simultaneous pervasiveness in nature of classically chaotic Hamiltonians and classical behavior is clearly perplexing. From a different viewpoint, however, it is the loss of coherence that occurs when a system interacts with its environment that permits a classical description of phenomena [3,4]. This line of reasoning, together with the main result of this paper, circumvents the difficulty raised above; we show that coherence decays faster in systems that are chaotic rather than regular. Hence, from this second viewpoint, the emergence of classical behavior should occur *sooner* in chaotic systems than in their regular counterparts.

Most investigations of decoherence and the emergence of classical behavior have relied on the simple model of the coupling of a system, described by Hamiltonian $H$, to a thermal environment of harmonic oscillators. The total Hamiltonian is taken as

$$H_T = H + \sum_j \hbar \omega_j b_j^\dagger b_j + \hbar A R, \quad (1)$$

where $R = \sum_j [\kappa(\omega_j) b_j + \kappa^*(\omega_j) b_j^\dagger]$, with $\kappa$ being a coupling parameter, and $A$ is a system operator. We wish to compare here the corresponding decoherence when $H$ is either classically regular or chaotic; only making use of generic properties of the two classes, we need not specify $H$ for our analysis. In the secular regime, we have previously calculated [5] the chaotic decay rates of coherences in the energy representation [6]. And, in the case of $A = H$, we found [7,8] that coherence, as measured by the purity $\text{Tr}\rho^2$, decays faster in chaotic systems than in their regular counterparts. Here we generalize some of these results and examine a more physical and much studied coupling, $A = q$, where $q$ is the position operator of the system. Recently [9], an analysis of a Brownian equation describing an unstable harmonic oscillator (potential$=-|k|q^2/2$) revealed that, at long times, the entropy of a Gaussian state increases linearly at a rate determined by $k$. In contrast, we use a different approach to look at generic regular and chaotic systems and examine not the long time entropy growth of a single Gaussian, but rather the decoherence of a superposition of two adjacent localized states. The perspective we adopt has been proposed previously [4,10,11] and may be summarized by the following excerpt [4]: "The destruction of interference terms is often considered as caused in a classical way by an 'uncontrollable influence' of the environment on the system of interest.... But the opposite is true: The system disturbs the environment, thereby dislocalizing the phases." We find that in some regimes chaotic systems disturb the environment more profoundly, resulting in faster decoherence.

It is instructive to first re-examine the case $A = H$ from this perspective; here the interaction with the environment can be thought of as performing quantum nondemolition measurements on the system's energy. Denote by $|E_n\rangle$ an energy eigenstate of the system and $|\phi_R\rangle$ an arbitrary reservoir state. An initial total operator $(|E_n\rangle\langle E_m|) \otimes (|\phi_R\rangle\langle\phi_R|)$ evolves exactly as

$$e^{-i\omega_{nm}t}(|E_n\rangle\langle E_m|) \otimes U_R(t, E_n)(|\phi_R\rangle\langle\phi_R|)U_R^\dagger(t, E_m), \quad (2)$$

where $\omega_{nm} \equiv (E_n - E_m)/\hbar$ and

$$U_R(t, E_n) = \exp\left\{-\frac{it}{\hbar}\left[H_R + \hbar E_n \sum_j \left(\kappa_j b_j + \kappa_j^* b_j^\dagger\right)\right]\right\}.$$

We see that the reservoir bra and ket are driven by different amplitudes if $n \neq m$. One would expect the reduced system operator [obtained by tracing expression (2) with respect to the reservoir variables] to decay faster, the greater the difference $|E_n - E_m|$. Since it is known [12] that nearest neighbor energy differences are greater in chaotic systems (the so called "energy repulsion") than in regular systems, our previous result [7,8] is further elucidated. Taking instead the dipole coupling $A = q$, we show below that a similar process in the Ehrenfest regime leads to faster decay of coherence in chaotic systems than in regular systems. What there takes the place of energy repulsion is "trajectory repulsion:" that is, the well-known exponentiation of chaotic trajectories.

## II. HARTREE ANALYSIS

We start our analysis with a two dimensional version of (1) with $A$ being replaced by the position $(q_x, q_y)$:

$$H_T = H + \sum_{\mathbf{j}} \sum_{n=1}^{2} \hbar \omega_{\mathbf{j}} b^\dagger_{\mathbf{j},n} b_{\mathbf{j},n} + \hbar q_x R_1 + \hbar q_y R_2, \quad (3)$$

where $R_{1(2)} = \sum_{\mathbf{j}}[\kappa(\omega_{\mathbf{j}})b_{\mathbf{j},1(2)} + \kappa^*(\omega_{\mathbf{j}})b^\dagger_{\mathbf{j},1(2)}]$.

Obtaining the exact evolution corresponding to $H_T$ for arbitrary $H$ is out of the question. We resort to an approximation in the spirit of a time-dependent, Hartree analysis [13]. Such an approach has been employed before to treat Hamiltonians similar to $H_T$ [14,11]. Graham and Höhnerbach [15] (see also [16] for related work) used this approximation to study a two level, chaotic system, although there a single mode treated classically (as opposed to a quantum reservoir) was coupled to the system; moreover, the chaos there was induced by strong coupling to the mode. This should be contrasted to our work below, where $H$ is intrinsically chaotic and the coupling to the reservoir is weak.

Suppose we take the following product initial state:

$$|\psi_{\text{in}}\rangle \otimes |\psi_{R,\text{in}}\rangle, \quad (4)$$

where $|\psi_{\text{in}}\rangle$ and $|\psi_{R,\text{in}}\rangle$ are, respectively, system and reservoir kets. We first find the best product state which approximates the evolution of (4) over a time long enough to capture the relevant decoherence; we are interested in the usual situation where this time is much shorter than that during which significant energy relaxation occurs. By "best" we mean that total state $|\psi_T(t)\rangle \equiv |\psi(t)\rangle \otimes |\psi_R(t)\rangle$, with initial condition (4), that results from the variational equation [17] $\langle\delta\psi_T|(i\hbar d/dt - H_T)|\psi_T\rangle = 0$. By considering variations in $\langle\psi|$ and $\langle\psi_R|$, the following coupled equations describing the evolution of $|\psi_T\rangle$ may be obtained:

$$i\hbar\frac{d|\psi\rangle}{dt} = \left( H + \frac{\langle\psi_R|R_1|\psi_R\rangle}{\langle\psi_R|\psi_R\rangle}\hbar q_x + \frac{\langle\psi_R|R_2|\psi_R\rangle}{\langle\psi_R|\psi_R\rangle}\hbar q_y \right.$$
$$\left. - \hbar\frac{\langle\psi|\otimes\langle\psi_R|(q_x R_1 + q_y R_2)|\psi_R\rangle\otimes|\psi\rangle}{\langle\psi_R|\psi_R\rangle\langle\psi|\psi\rangle} \right)|\psi\rangle,$$
(5)

$$i\hbar\frac{d|\psi_R\rangle}{dt} = \left( \sum_{\mathbf{j}} \sum_{n=1}^{2} \hbar \omega_{\mathbf{j}} b^\dagger_{\mathbf{j},n} b_{\mathbf{j},n} + \frac{\langle\psi|q_x|\psi\rangle}{\langle\psi|\psi\rangle}\hbar R_1 \right.$$
$$\left. + \frac{\langle\psi|q_y|\psi\rangle}{\langle\psi|\psi\rangle}\hbar R_2 \right)|\psi_R\rangle. \quad (6)$$

We call the solution of these last two equations a Hartree (total) ket and denote it by $|\psi_T(t)\rangle$. The error in this scheme is given by $|\Delta\psi_T(t)\rangle = |\psi_T^{\text{ex}}(t)\rangle - |\psi_T(t)\rangle$, where $|\psi_T^{\text{ex}}(t)\rangle$, is the exact solution of the Schrödinger equation with total Hamiltonian (3) and initial condition (4). By writing down the evolution equation that $|\Delta\psi_T(t)\rangle$ satisfies, switching to an interaction picture, and performing a Born expansion, a general expression for the error may be obtained to first order in the coupling parameter [11]; this we seek to characterize the initial Hartree kets $|\psi_{\text{in}}\rangle$ that lead to the most accurate description of early time behavior. Assuming a continuum of oscillator frequencies with density of modes $g(\omega)$, we adopt perhaps the most common model:

$$g(\omega)|\kappa(\omega)|^2 = \frac{C}{2\pi}\omega\Theta(\omega)\Theta(\omega_{\max} - \omega), \quad (7)$$

where $C$ is a coupling constant, $\Theta$ is the unit step function, and $\omega_{\max}$ is some high frequency cutoff. Noting that Eq. (6) evolves reservoir coherent states, denoted by $|\boldsymbol{\alpha}_R\rangle$ or $|\boldsymbol{\beta}_R\rangle$, into other reservoir coherent states, we choose them for the $|\psi_{R,\text{in}}\rangle$; they form a convenient basis with which to expand the reservoir canonical operator that we consider below [see state (9)], particularly so because they yield an asymptotic expression for the error that is independent of reservoir operators. To wit, if the coherent states are chosen for the $|\psi_{R,\text{in}}\rangle$, we find [18] that the error, to first order in the coupling, is then

$$\langle\Delta\psi_T(t)|\Delta\psi_T(t)\rangle \sim \frac{C}{2\pi}\overline{[(\Delta q_x)^2(t) + (\Delta q_y)^2(t)]} \quad (8)$$

as $\Omega_{\max} \equiv \omega_{\max}t \to \infty$ for a fixed time $t$, and where $\Delta q_{x(y)}$ is the uncertainty of the position $q_{x(y)}$ in the Hartree system ket $|\psi(t)\rangle$; the bar denotes the time average $t^{-1}\int_0^t [(\Delta q_x)^2(\tau) + (\Delta q_y)^2(\tau)] w_t(\tau)d\tau$ with weight

$$w_t(\tau) = \left\{ \frac{1 - \cos[\Omega_{\max}(1 - \tau/t)]}{1 - \tau/t} + \frac{1 - \cos(\Omega_{\max}\tau/t)}{\tau/t} \right\}.$$

For arbitrary $H$, it is a nontrivial task to determine which class of initial states $|\psi_{\text{in}}\rangle$ minimizes the error. If an eigenstate of $\mathbf{q}$ is chosen as the initial system ket, the uncertainty in the position, although initially zero, will quickly increase significantly because of the infinite momentum dispersion. For a nonzero time interval, relation (8) suggests that the smallest error results when the initial Hartree system ket $|\psi_{\text{in}}\rangle$ is a compromise between being completely localized in position or momentum. Only this qualitative feature of the $|\psi_{\text{in}}\rangle$, and its independence of the specific reservoir coherent state chosen for $|\psi_{R,\text{in}}\rangle$ under the approximation employed below, is indicated for the next step of our analysis.

As a simple example of a more general initial total state, in what follows we take the reservoir in thermal equilibrium and the system in a pure state consisting of a superposition of two such Hartree kets. That is, we consider the following initial condition:

$$\rho_{R,\text{can}} \otimes \sum_{i,j=1}^{2} c_i c_j^* |\mathbf{z}_i\rangle\langle\mathbf{z}_j|, \qquad (9)$$

where the Hartree system ket $|\mathbf{z}_j\rangle$ is centered about the four dimensional phase space point $\mathbf{z}_j$; $\rho_{R,\text{can}}$ is the reservoir operator corresponding to a canonical ensemble at temperature $T$. Now define $\chi(t)$ to be the trace, over the reservoir variables, of the operator that results by propagating $\rho_{R,\text{can}} \otimes |\mathbf{z}_1\rangle\langle\mathbf{z}_2|$ for a time $t$; the system operator $\chi(t)$ reflects the coherence that subsists between the two states $|\mathbf{z}_1\rangle$ and $|\mathbf{z}_2\rangle$. Determining the evolution of $\chi(t)$ entails that we solve Eqs. (5) and (6) self-consistently. A common technique to accomplish this is by iteration. One may start the iterative algorithm by assuming the evolution of the Hartree system ket to be unaffected by the reservoir so that $|\mathbf{z}_j^{(1)}(t)\rangle$, $j=1,2$, obeys

$$i\hbar\frac{d|\mathbf{z}_j^{(1)}(t)\rangle}{dt} = H|\mathbf{z}_j^{(1)}(t)\rangle, \qquad (10)$$

where the superscript of $|\mathbf{z}_j^{(n)}(t)\rangle$ refers to the $n$th iterative step. Next, the expectation value $\langle q_x\rangle(t,\mathbf{z}_j) \equiv \langle\mathbf{z}_j^{(1)}(t)|q_x|\mathbf{z}_j^{(1)}(t)\rangle$ and that for the $y$ coordinate are inserted into Eq. (6) and one then calculates $|\boldsymbol{\beta}_R^{(1)}(t)\rangle$. This is the end of one iteration and we may then form

$$\chi^{(1)}(t) = \text{Tr}_R \int \frac{d^2\boldsymbol{\beta}_R}{\pi} \int \frac{d^2\boldsymbol{\alpha}_R}{\pi} \langle\boldsymbol{\beta}_R|\rho_{R,\text{can}}|\boldsymbol{\alpha}_R\rangle$$
$$\times \left[\left(|\boldsymbol{\beta}_R^{(1)}(t)\rangle\langle\boldsymbol{\alpha}_R^{(1)}(t)|\right) \otimes \left(|\mathbf{z}_1^{(1)}(t)\rangle\langle\mathbf{z}_2^{(1)}(t)|\right)\right], \qquad (11)$$

where these last integrals run over the two polarizations and the infinite number of reservoir modes. For the second iteration, one would insert $\langle\boldsymbol{\beta}_R^{(1)}|b_j|\boldsymbol{\beta}_R^{(1)}\rangle$ ($j=1,2$), as well as the last mentioned system expectation values, into Eq. (5) and solve to obtain $|\mathbf{z}_j^{(2)}(t)\rangle$. Continuing in this manner, one could generate $\chi^{(2)}(t)$ and the higher iterates, the hope being that convergence to $\chi(t)$ occurs sufficiently rapidly. Even with this approximate scheme, calculating the $\chi^{(n)}(t)$ becomes increasingly prohibitive. Our simple goal here is to compute $\chi^{(1)}(t)$, leaving aside for now the difficult and important mathematical problem of determining the convergence properties of the $\chi^{(n)}(t)$ [19].

For a fixed time $t > 0$, we may calculate the controlling factor of the modulus of $\langle\mathbf{z}_1^{(1)}(t)|\chi^{(1)}(t)|\mathbf{z}_2^{(1)}(t)\rangle$. We find

$$\ln\left|\langle\mathbf{z}_1^{(1)}(t)|\chi^{(1)}(t)|\mathbf{z}_2^{(1)}(t)\rangle\right|$$
$$\sim -\frac{CkT}{2\hbar}\int_0^t \left\{[\langle q_x\rangle(\tau,\mathbf{z}_2) - \langle q_x\rangle(\tau,\mathbf{z}_1)]^2 + [\langle q_y\rangle(\tau,\mathbf{z}_2) - \langle q_y\rangle(\tau,\mathbf{z}_1)]^2\right\} d\tau \qquad (12)$$

as $\hbar\omega_{\max}/kT \to 0$ first, followed by $\Omega_{\max} \to \infty$ [20]. Henceforth, we shall restrict the system initial state by taking

$$\mathbf{z}_1 = \mathbf{z} \quad \text{and} \quad \mathbf{z}_2 = \mathbf{z} + \delta\mathbf{z} \qquad (13)$$

so that $|\mathbf{z}_2\rangle$ is now formally a state centered about a point in phase space only infinitesimally displaced from $\mathbf{z}_1$.

Consider now a system in the Ehrenfest regime [21], where the the expectation values $\langle\mathbf{q}\rangle(t)$ behave as would the classical coordinates $\mathbf{q}(t)$. We can then obtain universal behavior by appealing to the different rates of divergence of initially adjacent orbits. Generically [22], this divergence exhibits average linear [23,24] and exponential [25,24] growth; denoting by $\lambda > 0$ the maximum Lyapunov exponent [24] and by $\epsilon$ any positive number, we have

$$\int_0^t \left\{[q_x(\tau,\mathbf{z}+\delta\mathbf{z}) - q_x(\tau,\mathbf{z})]^2 \right.$$
$$\left. + [q_y(\tau,\mathbf{z}+\delta\mathbf{z}) - q_y(\tau,\mathbf{z})]^2\right\} d\tau$$
$$= \begin{cases} O(t^3), & \text{regular} \\ O(e^{2(1+\epsilon)\lambda t}), & \text{chaotic} \end{cases} \qquad (14)$$

as $t \to \infty$. In view of the corresponding expression in Eq. (12), for initial conditions (13) it is clear that coherence decays faster in the Ehrenfest regime for chaotic systems than for regular systems. The source of this faster decoherence is the unstable orbits. By imprinting their exponentiation on the reservoir states correlated with them, the coherence between them is lost at a correspondingly [Eqs. (12) and (14)] faster rate than it is for orbits in a regular system.

## III. DISCUSSION

For real, quantum systems, broaching this issue beyond the Ehrenfest regime is complicated by the fact that the "break time" $t_\hbar$ (i.e., the time after which the quantum expectation values are no longer approximated by the classical trajectories) is much shorter in chaotic than in regular systems. Although this matter is not completely resolved, some theoretical work [1,2] indicates that for chaotic and regular systems $t_\hbar$ is $O(\ln\hbar)$ and $O(\hbar)$, respectively (see also [26,27]). Nevertheless, numerical analyses [28] indicate that, for short times, narrow wave packets propagated quantum mechanically can exhibit large, classical-like sensitivities to initial conditions before spreading quickly dampens this instability. Moreover, because the differences of the expectation values in expression (12) enter as exponents, coherence decay will be greatly affected by any disparity in the divergence rates of neighboring wave packets—whether exponentially sensitive to initial conditions or not—in chaotic and regular systems.

As alluded to in the Introduction and discussed in [1], from one point of view the shorter break time in classically chaotic systems leads one to conclude that the greater the instability, the more a system resembles a quantum one. However, greater instability is *precisely* what gives rise to more effective decoherence when, unlike the previous point of view, the surroundings are taken into account. And when coherence is lost, the hallmark of quantum mechanics—the interference effects that arise from the quantum prescription of adding amplitudes, and not probabilities, of alternative events—is quashed. By demonstrating that orbital instability leads to faster decoherence, we have thus established an important link between chaos and the appearance of classical behavior.

---

[1] G.M. Zaslavsky, in *Quantum Dynamics of Chaotic Systems*, edited by J.-M. Yuan, D.H. Feng, and G.M. Zaslavsky (Gordon and Breach, Langhorne, PA, 1993).
[2] M.V. Berry and N.L. Balazs, J. Phys. A **12**, 625 (1979).
[3] W.H. Zurek, Phys. Rev. D **24**, 1516 (1981); **26**, 1862 (1982); Phys. Today **44** (10), 36 (1991); C.M. Savage and D.F. Walls, Phys. Rev. A **32**, 2316 (1985); J.E. Sipe and N. Arkani-Hamed, *ibid.* **46**, 2317 (1992).
[4] E. Joos and H.D. Zeh, Z. Phys. B **59**, 223 (1985).
[5] A. Tameshtit and J.E. Sipe, Phys. Rev. A **49**, 89 (1994).
[6] For a temporally periodic perturbation applied to a Hamiltonian system (typically having one degree of freedom) in the absence of any thermal effects, Berman and Zaslavsky [29] found semiclassically that the off-diagonal elements of the density operator (in the energy representation of the unperturbed Hamiltonian) decay *only if* the perturbation induces chaos. Again, neglecting thermal noise, but including dissipation, Dittrich and Graham [30] examined the stroboscopically kicked (and hence nonconservative) rotor only within the rotating wave approximation; restricting this specific model to the chaotic diffusive regime and using order of magnitude estimates, they found that the phase coherence lifetime of the quasienergies was inversely proportional to the square of the kicking strength. See also Cohen [31], for further work on the kicked rotor or particle, and Blumel *et al.* [32].
[7] A. Tameshtit and J.E. Sipe, in *Quantum Dynamics of Chaotic Systems* [1].
[8] A. Tameshtit and J.E. Sipe, Phys. Rev. A **47**, 1697 (1993).
[9] W.H. Zurek and J.P. Paz, Phys. Rev. Lett. **72**, 2508 (1994).
[10] A. Stern, Y. Aharonov, and Y. Imry, Phys. Rev. A **41**, 3436 (1990).
[11] N. Arkani-Hamed and J.E. Sipe (unpublished).
[12] F. Haake, *Quantum Chaos* (Springer-Verlag, Berlin, 1991), and references cited therein.
[13] D.R. Hartree, *The Calculation of Atomic Structures* (Wiley, New York, 1957).
[14] H. Primas, in *Sixty-Two Years of Uncertainty*, edited by A.I. Miller (Plenum Press, New York, 1990), p. 259.
[15] R. Graham and M. Höhnerbach, Z. Phys. B **57**, 233 (1984).
[16] L.L. Bonilla and F. Guinea, Phys. Lett. B **271**, 196 (1991); Phys. Rev. A **45**, 7718 (1992).
[17] J.I. Frenkel, *Wave Mechanics, Part II* (Oxford University Press, Oxford, 1934).
[18] Some technical assumptions were used to derive (8) from the first order expression; for example, we assumed $\langle\psi(t)|q_x|\psi(t)\rangle$ and the analogous expression for $y$ to be a thrice continuously differentiable function of time. Note also that these position expectation values depend on the cutoff frequency $\omega_{\max}$.
[19] We have not investigated to what extent this Hartree approximation captures the initial slip or jolt [33] previously found in analyses of a system harmonic oscillator coupled to a bath of harmonic oscillators. Inasmuch as the jolt is partly due to unphysical initial conditions, and appears to have no significant effect on decoherence when the coupling is weak [34], this is probably not a serious drawback.
[20] See also [10] for related expressions involving the interference of two given electron paths (that start and end together) in a metal.
[21] For a discussion of the validity of the Ehrenfest approximation see, e.g., A. Messiah, *Quantum Mechanics* (North-Holland, Amsterdam, 1958), Vol. 1, Chap. VI.
[22] Harmonic oscillators and periodic orbits are excluded.
[23] G. Casati, B.V. Chirikov, and J. Ford, Phys. Lett. **77A**, 91 (1980).
[24] H.-D. Meyer, J. Chem. Phys. **84**, 3147 (1986).
[25] V.I. Oseledec, Trans. Moscow Math. Soc. **19**, 197 (1968).
[26] R.G. Littlejohn, Phys. Rep. **138**, 193 (1986).
[27] S. Tomsovic and E.J. Heller, Phys. Rev. E **47**, 282 (1993).
[28] R.S. Judson, S. Shi, and H. Rabitz, J. Chem. Phys. **90**, 2274 (1989).
[29] G.P. Berman and G.M. Zaslavsky, Physica **97A**, 367 (1979); G.M. Zaslavsky, *Chaos in Dynamic Systems* (Harwood, Chur, Switzerland, 1985), Chap. 11.
[30] T. Dittrich and R. Graham, Ann. Phys. **200**, 363 (1990).
[31] D. Cohen, Phys. Rev. A **43**, 639 (1991); **44**, 2292 (1991).
[32] R. Blumel, A. Buchleitner, R. Graham, L. Sirko, U. Smilansky, and H. Walther, Phys. Rev. A **44**, 4521 (1991).
[33] F. Haake and R. Reibold, Phys. Rev. A **32**, 2462 (1985).
[34] J.P. Paz, S. Habib, and W.H. Zurek, Phys. Rev. D **47**, 488 (1993).